\documentstyle[preprint,aps,eqsecnum]{revtex}

\newcommand{\lbar}{\overline}
\newcommand{\spa}{\vspace{.25cm}}
\newcommand{\beq}{\begin{equation}}
\newcommand{\eeq}{\end{equation}}
\newcommand{\beqs}{\begin{eqnarray}}
\newcommand{\eeqs}{\end{eqnarray}}

\newcommand{\BR}{\mbox{{\rm BR}}}
\newcommand{\FCNS}{$\nu_\mu \, f \to \nu_\tau \, f$}
\newcommand{\TauToMuee}{\tau^- \to \mu^- \,e^+ \,e^-}
\newcommand{\TauToMuM}{\tau^- \to \mu^- \, M~(M=\pi^0,\rho^0,\eta)}
\newcommand{\TauToMuRho}{\tau^- \to \mu^- \, \rho^0}
\newcommand{\TauToMuPi} {\tau^- \to \mu^- \, \pi^0}
\newcommand{\TauToMuEta}{\tau^- \to \mu^- \, \eta}
\newcommand{\TauToMuPiPi}{\tau^- \to \mu^- \, \pi^+ \pi^-}
\newcommand{\sg}{{\bf 1}}
\newcommand{\db}{{\bf 2}}
\newcommand{\tb}{{\bf 3}}
\newcommand{\ta}{\bar \tb}

\newcommand{\SUSYwoR}{SUSY $\not\!\!R_p$}
\newcommand{\calS}{{\cal S}}
\newcommand{\calV}{{\cal V}}
\newcommand{\epj}[2]{{\em Eur. Phys. J.} {\bf C\,#1}, #2 }
\newcommand{\sci}[2]{{\em Science} {\bf  #1}, #2 }

\def\lsim{\ \rlap{\raise 3pt \hbox{$<$}}{\lower 3pt \hbox{$\sim$}}\ }
\def\gsim{\ \rlap{\raise 3pt \hbox{$>$}}{\lower 3pt \hbox{$\sim$}}\ }

\def\npb#1{Nucl.\ Phys.\ {\bf B\,#1}}
\def\plb#1{Phys.\ Lett.\ {\bf B\,#1}}
\def\prd#1{Phys.\ Rev.\ {\bf D\,#1}}
\def\prl#1{Phys.\ Rev.\ Lett. {\bf#1}}

\def\zpc#1{Z.~Phys.\ {\bf C\,#1}}
\def\epjc#1{Eur.~Phys.~J.\ {\bf C\,#1}}

\def\mpla#1{Mod. Phys. Lett. {\bf A\,#1}}
\def\JHEP#1{JHEP {\bf #1}}
\def\col{collab.}

\begin{document}
\draft
{\tighten 
\preprint{\vbox{\hbox{WIS-99/31/Sept-DPP}
                \hbox{SLAC-PUB-8244}
                \hbox{hep-ph/9909390}
                \hbox{September 1999}}}
\title{~ \\ Can lepton flavor violating interactions \\ 
            explain the atmospheric neutrino problem?}
\footnotetext{\scriptsize Research at SLAC is supported by the 
                          U.S. Department of Energy 
                          under contract DE-AC03-76SF00515.}
\author{Sven Bergmann,\,$^a$ Yuval Grossman\,$^b$ 
        and Damien M. Pierce\,$^c$}
\address{ \vbox{\vskip 0.truecm}
  $^a$Department of Particle Physics \\
  Weizmann Institute of Science, Rehovot 76100, Israel \\
\vbox{\vskip 0.truecm}
  $^b$Stanford Linear Accelerator Center \\
  Stanford University, Stanford, CA 94309\\
\vbox{\vskip 0.truecm}
  $^c$Brookhaven National Laboratory\\
  Upton, NY 11973}

\maketitle

\begin{abstract}
 
\noindent
We investigate whether flavor changing neutrino interactions (FCNIs)
can be sufficiently large to provide a viable solution to the
atmospheric neutrino problem. Effective operators induced by heavy
boson exchange that allow for flavor changing neutrino scattering off
quarks or electrons are related by an $SU(2)_L$ rotation to operators
that induce anomalous tau decays.  Since $SU(2)_L$ violation is small
for New Physics at or above the weak scale, one can use the upper
bounds on lepton flavor violating tau decays or on lepton universality
violation to put severe, model-independent bounds on the relevant
non-standard neutrino interactions. Also $Z$-induced flavor changing
neutral currents, due to heavy singlet neutrinos, are too small to be
relevant for the atmospheric neutrino anomaly. We conclude that the
FCNI solution to the atmospheric neutrino problem is ruled out.

\end{abstract} 
} 

\newpage

\section{Introduction}

Several atmospheric neutrino (AN) experiments~\cite{KK,IMB,Soudan,SK}
observe an anomalous ratio $\nu_e/\nu_\mu$ in the atmospheric neutrino
flux. This long-standing AN problem has been confirmed by the recent
Super-Kamiokande high-statistics observations~\cite{SK}, which give
strong evidence that the standard model (SM) description of the
neutrino sector is incomplete. The standard solution to the AN anomaly
in terms of neutrino oscillations requires that neutrinos are massive,
and that there is mixing in the lepton sector. Then, $\nu_\mu \to
\nu_\tau$ oscillations can explain the atmospheric neutrino data
provided that the relevant mass-squared difference is $\Delta_{AN}
\sim 10^{-3} ~\mbox{eV}^2$ and the muon and tau neutrinos have large
vacuum mixing angles, $\sin 2\theta_{AN} \sim 1$.

Recently, an alternative solution, where the AN anomaly is induced by
non-standard neutrino interactions has been
proposed~\cite{MaRoy,Brooijmans,GGea,Fornengo}. In this scenario the
neutrinos are assumed to be massless, but they are subject to
non-standard interactions.  For neutrino propagation in matter, flavor
changing neutrino interactions (FCNIs) induce an off-diagonal term in
the effective neutrino mass matrix, while non-universal flavor
diagonal interactions generate the required splitting between the
diagonal terms. A priori such a scenario is well motivated, since many
extension of the SM predict new neutrino interactions. Moreover, it is
well known that one cannot explain the atmospheric~\cite{AN},
solar~\cite{SN} and LSND~\cite{LSND} neutrino anomalies with three
light neutrinos. Thus, rather than ignoring one of the results or
introducing a forth, light sterile-neutrino~\cite{sterile}, it is
interesting to investigate whether FCNIs can explain any of the three
neutrino anomalies~\cite{BergmannGrossman}.

The two effective parameters that describe the non-standard
interactions of $\nu_\mu$ and $\nu_\tau$
are~\cite{Roulet,Barger91b,Brooijmans,GGea,Fornengo}
\beq \label{defeps}
\epsilon_\nu^f \equiv 
 {G_{\nu_\mu \nu_\tau}^f \over G_F}~~~~\mbox{and}~~~~ 
{\epsilon'}_\nu^f \equiv 
 {G_{\nu_\tau \nu_\tau}^f - G_{\nu_\mu \nu_\mu}^f \over G_F} \,,
\eeq
where $G_{\nu_\alpha \nu_\beta}^f$ ($\alpha, \beta =\mu, \tau$ and
$f=u,d,e$) denotes the effective coupling of the four fermion
operator
\beq \label{Onu}
{\cal O}_\nu^f \equiv (\lbar{\nu_\alpha} \, \nu_\beta) \, (\bar f \, f) \,.
\eeq
The Lorentz structure of ${\cal O}_\nu^f$ depends on the New Physics
that induces this operator. Operators which involve only left-handed
neutrinos (and which conserve total lepton number) can be decomposed
into a $(V-A) \otimes (V-A)$ and a $(V-A) \otimes (V+A)$ component.
(Any single New Physics contribution that is induced by chiral
interactions yields only one of these two components.) It is, however,
important to note that only the vector part of the background fermion
current affects the neutrino propagation for an unpolarized medium at
rest~\cite{BGN}. Hence only the $(V-A) \otimes (V)$ part of ${\cal
  O}_\nu^f$ is relevant for neutrino oscillations in normal matter.
One mechanism to induce such operators is due to the exchange of heavy
bosons that appear in various extensions of the standard model.  An
alternative mechanism arises when extending the fermionic sector of
the standard model and is due to $Z$-induced flavor-changing neutral
currents (FCNCs)~\cite{Z-FCNC,BergmannKagan}.

Recent analyses~\cite{Brooijmans,GGea,Fornengo} of non-standard
neutrino interactions as a possible solution to the atmospheric
neutrino data suggest that FCNIs can provide a good fit to the data
provided that
\beq \label{epsrequired}
\epsilon_\nu^q, {\epsilon'}_\nu^q \gsim 0.1
\quad {\rm or} \quad
\epsilon_\nu^e, {\epsilon'}_\nu^e \gsim 0.3 \,.
\eeq
In Ref.~\cite{Brooijmans,GGea,Fornengo} only new interactions
involving the $d$ quark were considered. Since the earth is
electrically neutral and its neutron to proton ratio is close to
unity, we conclude that the required values for $\epsilon_\nu^q$ and
${\epsilon'}_\nu^q$ are similar for $q=d,u$, while those for
$\epsilon_\nu^e$ and ${\epsilon'}_\nu^e$ are larger by a factor of
three.

The authors of Ref.~\cite{Lipari,Fogli-AN} argue that such a scenario
does not lead to a good description of the data.  In this paper we do
not try to resolve this issue, but investigate whether the lower
bounds on $\epsilon_\nu^f$ and ${\epsilon'}_\nu^f$
in~(\ref{epsrequired}) are at all phenomenologically viable. The
authors of Ref.~\cite{Fornengo} have estimated the upper bound
$\epsilon_\nu^d \lsim 0.1-1.0$ from the low energy $\nu_\mu$ neutral
current cross-section, concluding that FCNIs could be relevant for the
AN anomaly. In Ref.~\cite{Guzzo} specific models that could give
$\epsilon_\nu^f$ and ${\epsilon'}_\nu^f$ as large as
in~(\ref{epsrequired}) were discussed. We argue, however, that
model-independently the upper bounds from related, charged lepton
decay data imply that $\epsilon_\nu^f$ or ${\epsilon'}_\nu^f$ can be
at most at the one-percent level.  Thus we conclude that FCNIs do not
play a significant role for the atmospheric neutrino problem.

In Section~\ref{newbosons} we investigate in a model-independent
framework the constraints on FCNIs that are induced by heavy boson
exchange.  In most cases the upper bounds on lepton flavor violating
tau decays, in particular $\TauToMuM$ and $\TauToMuee$, imply
stringent constraints on $\epsilon_\nu^f$ that are inconsistent
with~(\ref{epsrequired}).  In the remaining cases severe constraints
on ${\epsilon'}_\nu^f$ are derived using bounds on lepton universality
violation.  In Section~\ref{Z-FCNC} we show that also $Z$-induced
FCNCs, that arise due to heavy singlet neutrinos, cannot be large
enough to explain the AN anomaly. We conclude in
Section~\ref{conclusions}.

\section{Flavor changing neutrino interactions induced 
         by heavy boson exchange}
\label{newbosons}

\subsection{Formalism}

The analysis of FCNIs that could be relevant for the AN problem is
similar to the discussion in Ref.~\cite{BergmannGrossman}, where the
possibility that FCNIs explain the LSND results~\cite{LSND} was ruled
out. In general, the presence of a heavy boson ${\cal B}$ that couples
to fermion bilinears $B_{ij}$ with the trilinear couplings
$\lambda_{ij}$, where $i, j = 1, 2, 3$ refer to fermion generations,
gives rise to the four fermion operator $B_{ij}^\dagger B_{kl}$ with
the effective coupling
\beq \label{GN} 
G^{B^\dagger B}_N = {{\lambda_{ij}^*} \lambda_{kl} \over 
                     {4 \sqrt2 M^2_{\cal B} } } \,, 
\eeq
at energies well below the boson mass $M_{\cal B}$.  Thus, in terms of
the trilinear coupling $\lambda_{\alpha f}$ that describes the
coupling of some heavy boson ${\cal B}$ to $\nu_\alpha$
($\alpha=\mu,\tau$) and a charged fermion $f=u,d,e$ the effective
parameters in~(\ref{defeps}) are given by
\beq \label{geneps}
\epsilon_\nu^f =    {\lambda_{\tau f}^* \lambda_{\mu f} 
                     \over  4 \sqrt2 M^2_{\cal B} \, G_F }
~~~~\mbox{and}~~~~ 
{\epsilon'}_\nu^f = {|\lambda_{\tau f}|^2 - |\lambda_{\mu f}|^2 
                     \over  4 \sqrt2 M^2_{\cal B} \, G_F } \,.
\eeq
The crucial point of our analysis is the following: Since the SM
neutrinos are components of $SU(2)_L$ doublets, the same trilinear
couplings $\lambda_{\alpha f}$ that give rise to non-zero
$\epsilon_\nu^f$ or ${\epsilon'}_\nu^f$ also induce other four-fermion
operators.  These operators involve the $SU(2)_L$ partners of the
neutrinos, i.e.  the charged leptons, and can be used to constrain the
relevant couplings.

\spa
\begin{center} 
\begin{tabular}{| c || c || c || c |} 
\hline  
\label{tab1}
~Bilinear $B$~   & ~Couples to Boson ${\cal B}$~ & Example 
                 & $(M_1/M_2)^2_{\rm max}$ \cr 
\hline \hline  
$(L L)_s$        & $\calS(\sg,\sg,1)$ & $~\tilde \ell^c_R$ (\SUSYwoR)~ & \cr
\hline
$\bar L \ell_R$  & $\calS(\sg,\db,1/2)$ & $\tilde L^c$ (\SUSYwoR) & 6.8 \cr
\hline
$(L L)_t$        & $\calS(\sg,\tb,1)$ & $\Delta_L$ (LRSM) & 5.9 \cr
\hline \hline
$(\bar L L)_s$   & $\calV(\sg,\sg,0)$ & & \cr
\hline
$L \ell_R$       & $\calV(\sg,\db,3/2)$ & & \cr
\hline
$(\bar L L)_t$   & $\calV(\sg,\tb,0)$ & & \cr
\hline \hline
$\lbar{e_R} e_R$ & $\calV(\sg,\sg,0)$ & & \cr
\hline
\end{tabular} 

\spa
Tab.~1: Lepton-Lepton Bilinears \\
\end{center}

\begin{center} 
\begin{tabular}{| c || c || c || c |} 
\hline  
\label{tab2}
~Bilinear $B$~  & ~Couples to Boson ${\cal B}$~ 
                & Example &  $(M_1/M_2)^2_{\rm max}$ \cr
\hline \hline  
$(L Q)_s$       & $\calS(\ta,\sg,1/3)$ & $~\tilde d^c_R$ (\SUSYwoR)~ & \cr
\hline
$\bar L d_R$    & $\calS(\ta,\db,-1/6)$ & $\tilde Q^c$ (\SUSYwoR) & 5.2 \cr
\hline
$\bar L u_R$    & $\calS(\ta,\db,-7/6)$ & & 3.6 \cr
\hline
$(L Q)_t$       & $\calS(\ta,\tb,1/3)$ & & 2.5 \cr
\hline \hline
$(\bar L Q)_s$  & $\calV(\ta,\sg,-2/3)$ & & \cr
\hline
$L d_R$         & $\calV(\ta,\db,5/6)$ & & \cr
\hline
$L u_R$         & $\calV(\ta,\db,-1/6)$ & & \cr
\hline
$(\bar L Q)_t$  & $\calV(\ta,\tb,-2/3)$ & & \cr
\hline
\end{tabular} 

\spa
Tab.~2: Lepton-Quark Bilinears \\
\end{center}

\begin{center} 
\begin{tabular}{| c || c |} 
\hline  
\label{tab3}
~Bilinear $B$~   & ~Couples to Boson ${\cal B}$~  \cr
\hline \hline  
$(\bar Q Q)_s$   & $\calV(\sg,\sg,0)$ \cr
\hline
$\lbar{u_R} u_R$ & $\calV(\sg,\sg,0)$ \cr
\hline
$\lbar{d_R} d_R$ & $\calV(\sg,\sg,0)$ \cr
\hline
$(\bar Q Q)_t$   & $\calV(\sg,\tb,0)$ \cr
\hline
\end{tabular} 

\spa
Tab.~3: Quark-Quark Bilinears \\
\end{center}

In order to obtain a complete list of these operators we note that
Lorentz invariance implies that any fermionic bilinear $B_{ij}$ can
couple to either a scalar~($\calS$) or a vector~($\calV$) boson.  If
the two fermions of the bilinear have the same (opposite) chirality
they require scalar (vector) couplings. To form a gauge invariant
trilinear coupling, the boson~${\cal B}$ must have opposite
hypercharge $Y$ and transform in the appropriate representation of
$SU(2)_L$ and $SU(3)_C$. Since the SM neutrinos only appear in
doublets of $SU(2)_L$ and since all right-handed (left-handed) charged
fermions transform as $SU(2)_L$ singlets (doublets), it follows that
any boson ${\cal B}$ that couples to the fermionic bilinear can only
be a singlet~($s$), a doublet~($d$) or a triplet~($t$) of $SU(2)_L$.
All relevant bilinears containing only leptons are listed in Tab.~1,
and those that are built from a lepton doublet and a quark are listed
in Tab.~2. In Tab.~3 we list the relevant diquark bilinears, namely,
those that can couple to $\bar L L$. Here $Q$ and $L$ denote the
left-handed SM quark and lepton doublets, and $e_R, u_R, d_R$ refer to
the right-handed SM singlets. Some of these couplings appear in well
known extensions of the standard model. For example, in supersymmetric
models without $R$-parity (\SUSYwoR)~\cite{SUSY}, fermion bilinears
can couple to left-handed sleptons ($\tilde L^c$), right-handed
sleptons ($\tilde \ell^c_R$), left-handed squarks ($\tilde Q^c$) and
right-handed down squarks ($\tilde d^c_R$), as indicated in the third
column of the tables. An example for a scalar triplet is the
$\Delta_L$ in left-right symmetric models (LRSMs)~\cite{LRSM}.

In general any two bilinears appearing in Tab.~1--3 that couple to the
same boson can be combined to a four fermion interaction with
effective coupling as given in Eq.~(\ref{GN}). In order to generate a
non-zero $\epsilon_\nu^f$ or ${\epsilon'}_\nu^f$ in Eq.~(\ref{geneps})
at least one of the bilinears has to contain a lepton doublet $L$.
Clearly, four fermion operators that are the product of one bilinear
and its hermitian conjugate can be constructed. If the two bilinears
have the same (different) flavor structure the resulting operator will
conserve (violate) lepton flavor.  In addition, the $(\bar L L)$
bilinear can couple to $(\lbar{e_R} e_R)$ or to any of the quark-quark
bilinears in Tab.~3 inducing four-Fermi interactions of the form
$(\bar L L)(\lbar{e_R} e_R)$, $(\bar L L)(\bar Q Q)$, $(\bar L
L)(\lbar{u_R} u_R)$ and $(\bar L L)(\lbar{d_R} d_R)$.

Note that scalar or vector fields that transform identically under the
unbroken $SU(3)_C \otimes U(1)_{EM}$ symmetry can mix. If this mixing
is between a doublet and a singlet or a triplet the resulting
operators violate total lepton number and are not relevant for our
analysis. In the case of singlet--triplet mixing no new operators are
generated. Therefore this kind of mixing does not affect our
conclusions and we neglect it.

To demonstrate how $SU(2)_L$ related processes can be used to
constrain the parameters $\epsilon_\nu^f$ or ${\epsilon'}_\nu^f$, let
us consider for example the bilinear $\bar L f_R~(f=e,u,d)$ that
couples via a scalar doublet to its hermitian conjugate $\lbar{f_R}
L$. In terms of the component fields the effective interaction is
\beqs
&~&
{\lambda^*_{\alpha f} \lambda_{\beta f} \over M_1^2}
 (\lbar{\nu_\alpha} f_R)\,(\lbar{f_R} \nu_\beta) + 
{\lambda^*_{\alpha f} \lambda_{\beta f} \over M_2^2}
 (\lbar{\ell_\alpha} f_R)\,(\lbar{f_R} \ell_\beta) \nonumber \\
&=& \label{fierzed}
-{\lambda^*_{\alpha f} \lambda_{\beta f} \over 2 M_1^2}
  (\lbar{\nu_\alpha} \gamma^\mu \nu_\beta)\,(\lbar{f_R} \gamma_\mu f_R)  
-{\lambda^*_{\alpha f} \lambda_{\beta f} \over 2 M_2^2}
  (\lbar{\ell_\alpha} \gamma^\mu \ell_\beta)\,(\lbar{f_R} \gamma_\mu f_R) 
\,, 
\eeqs
where $\ell_\alpha = \mu_L, \tau_L$ for $\alpha = \mu, \tau$.
$\lambda_{\alpha f}$ is the trilinear coupling of $\lbar{L_\alpha}
f_R$ to the scalar doublet and $M_{1,2}$ denote the masses of its
$SU(2)_L$ components. The important point is that the scalar
doublet exchange not only gives rise to the four-Fermi operator 
${\cal O}^f_\nu$ in~(\ref{Onu}) (with $(V-A) \otimes (V+A)$ 
structure), but also produces the $SU(2)_L$ related operator
\beq \label{Oell}
{\cal O}_\ell^f \equiv 
 (\lbar{\ell_\alpha} \, \ell_\beta) \, (\bar f \, f) \,,
\eeq
which has the same Lorentz structure as ${\cal O}_\nu^f$, but where
the neutrinos are replaced by their charged lepton partners. Moreover,
the effective coupling of ${\cal O}^f_\ell$, that we denote by
$G_{\alpha \beta}^f$, is related to $G_{\nu_\alpha \nu_\beta}^f$ by
\beq \label{Gratio}
G_{\nu_\alpha \nu_\beta}^f = G_{\alpha \beta}^f \, {M_1^2 \over M_2^2}\,.
\eeq

Constructing all the relevant four fermion operators that are induced
by the couplings between the bilinears listed in Tab.~1--3, one finds
that in general ${\cal O}^f_\ell$ is generated together with ${\cal
O}^{f'}_\nu$. Here $f'$ can be different from $f$ only for
interactions with quarks, that is in some cases ${\cal O}^u_\ell$
(${\cal O}^d_\ell$) is generated together with ${\cal O}^d_\nu$
(${\cal O}^u_\nu$). The leptonic operator ${\cal O}^e_\ell$ is always
generated together with ${\cal O}^e_\nu$ unless the interaction is
mediated by an intermediate scalar $SU(2)_L$ singlet that couples to
\beq
(L_\alpha L_e)_s = 
{1 \over \sqrt2} (\lbar{\nu_\alpha^c} e_L - \lbar{\ell_\alpha^c} \nu_e) \,,
\eeq
where $\ell_\alpha = \mu_L, \tau_L$ for $\alpha=\mu, \tau$, with the
elementary coupling $\lambda_{\alpha e}$.  The coupling of $(L_\alpha
L_e)_s$ to $(L_\beta L_e)_s^\dagger$ that is mediated by a scalar
singlet of mass $M$ yields the effective interactions
\beqs
&~&
{\lambda^*_{\beta e} \lambda_{\alpha e} \over M^2} 
\left[
 (\lbar{e_L} \nu_\beta^c)\,(\lbar{\nu_\alpha^c} e_L) 
-(\lbar{e_L} \nu_\beta^c)\,(\lbar{\ell_\alpha^c} \nu_e) 
+(\lbar{\nu_e} \ell_\beta^c)\,(\lbar{\ell_\alpha^c} \nu_e) 
-(\lbar{\nu_e} \ell_\beta^c)\,(\lbar{\nu_\alpha^c} e_L) 
\right] = \\
&~& 
{\lambda^*_{\beta e} \lambda_{\alpha e} \over 2 M^2} 
\left[
 (\lbar{e_L} \gamma^\mu e_L)\,(\lbar{\nu_\beta} \gamma_\mu \nu_\alpha)
-(\lbar{e_L} \gamma^\mu \nu_e)\,(\lbar{\nu_\beta} \gamma_\mu \ell_\alpha) 
+(\lbar{\nu_e} \gamma^\mu \nu_e)\,(\lbar{\ell_\beta} \gamma_\mu \ell_\alpha)
-(\lbar{\nu_e} \gamma^\mu e_L)\,(\lbar{\ell_\beta} \gamma_\mu \nu_\alpha)
\right] \,, \nonumber \\
\label{fierzed2}
\eeqs
where we used a Fierz transformation and the identity $\lbar{A^c}
\gamma^\mu B^c = -\lbar{B} \gamma^\mu A$ to obtain~(\ref{fierzed2}).
One can see that in this case ${\cal O}^e_\nu$ is generated together
with three more operators that have the same effective coupling (up to
a sign). However, unlike for the case of intermediate doublets or
triplets, all these operators involve two charged leptons and two
neutrinos.


\subsection{Experimental constraints}

There is no experimental evidence for any non-vanishing $G_{\mu
\tau}^f$\,. Therefore, whenever ${\cal O}^f_\ell$ is generated
together with ${\cal O}^f_\nu$, one can use the upper bounds on
$G_{\mu \tau}^f$ to derive constraints on $G_{\nu_\mu \nu_\tau}^f$.
The most stringent constraint on $G_{\mu \tau}^e$ is due to the upper
bound on $\TauToMuee$~\cite{Bliss,PDG}:
\beq
\BR(\TauToMuee) ~<~ 1.7 \times 10^{-6} \,. \label{eeBound} \\
\eeq
Normalizing the above bound to the measured rate of the related lepton
flavor conserving decay, $\BR(\tau^- \to \mu^- \, \lbar{\nu_\mu} \,
\nu_\tau) = 0.17$~\cite{PDG}, we obtain
\beq \label{GeBound}
G_{\mu \tau}^e < 3.1 \times 10^{-3} \, G_F \,.
\eeq

To constrain $G_{\mu \tau}^q$ we may use the upper bounds on various
semi-hadronic tau decays that violate lepton flavor~\cite{Bliss,PDG}:
\beqs
\BR(\TauToMuPi)   ~&<&~ 4.0 \times 10^{-6} \,, \label{PiBound}  \\
\BR(\TauToMuRho)  ~&<&~ 6.3 \times 10^{-6} \,, \label{RhoBound} \\
\BR(\TauToMuEta)  ~&<&~ 9.6 \times 10^{-6} \,, \label{EtaBound} \\
\BR(\TauToMuPiPi) ~&<&~ 8.2 \times 10^{-6} \,. \label{PiPiBound} 
\eeqs
Let us first consider the tau decays into $\pi^0$ and $\rho^0$.  Since
these mesons belong to an isospin triplet we can use the isospin
symmetry to normalize the above bounds~(\ref{PiBound})
and~(\ref{RhoBound}) by the measured rates of related lepton flavor
conserving decays.  Using $\BR(\tau^- \to \nu_\tau \pi^-) =
0.11$~\cite{PDG} and $\BR(\tau^- \to \nu_\tau \rho^-) =
0.22$~\cite{Don,PDG} we obtain
\beq \label{GqBound}
G_{\mu \tau}^q(\pi)  < 8.5 \times 10^{-3} \, G_F~~~~\mbox{and}~~~~
G_{\mu \tau}^q(\rho) < 7.5 \times 10^{-3} \, G_F \,. 
\eeq
Since the $\pi$ ($\rho$) is a pseudoscalar (vector) meson
its decay probes the axial-vector (vector) part of the quark current.

In general, any semi-hadronic operator ${\cal O}^q_\ell$ can be
decomposed into an $I=0$ and an $I=1$ isospin component.  Only the
effective coupling of the latter can be constrained by the upper
bounds on the decays into final states with isovector mesons, like the
$\pi$ and the $\rho$.  If the resulting operator is dominated by the
$I=0$ component, the bounds in~(\ref{GqBound}) do not hold. But in
this case we can use the upper bound on $\BR(\TauToMuEta)$
in~(\ref{EtaBound}).  Since the $\eta$ is an isosinglet, isospin
symmetry is of no use for the normalization.  However, we can estimate
the proper normalization using the relation between the $\eta$ and
$\pi$ hadronic matrix elements, which is just the ratio of the
respective decay constants, $f_\eta/f_\pi \simeq 1.3$~\cite{Don,PDG}.
Taking into account the phase space effects, we obtain
from~(\ref{EtaBound}) that
\beq \label{GqBoundEta}
G_{\mu \tau}^q(\eta) < 1.2 \times 10^{-2} \, G_F \,.
\eeq
Since the $\eta$ is a pseudoscalar meson its decay probes the
axial-vector part of the $I=0$ component of the quark current, while
the neutrino propagation is only affected by the vector part.  As we
have already mentioned, for any single chiral New Physics contribution
the vector and axial-vector parts have the same magnitude and we can
use~(\ref{GqBoundEta}) to constrain the isosinglet component of ${\cal
O}_\ell^q$. In case there are several contributions, whose
axial-vector parts cancel each other (a vector singlet
$\calV(\sg,\sg,0)$ that couples to all the diquark singlets of Tab.~3
with the same strength would lead to such a scenario), the $I=0$
component could still be constrained by the upper bound on
$\BR(\TauToMuPiPi)$ in~(\ref{PiPiBound}). While the calculation of the
rate is uncertain due to our ignorance of the spectra and the decay
constants of the isosinglet scalar resonances, we expect that the
normalization will be similar to that of the $\pi$, $\rho$ and $\eta$
discussed before. Finally we note that the decay $\tau^- \to \mu^- \,
\omega$ would be ideal to constrain the $I=0$ vector part, but at
present no upper bound on its rate is available.

While one can always fine-tune some parameters in order to avoid our
bounds, our basic assumption is that this is not the case.  Thus
from~(\ref{GeBound}), (\ref{GqBound}) and~(\ref{GqBoundEta}) we
conclude that the effective coupling $G_{\mu \tau}^f$ could be at most
at the one-percent level.

We still have to discuss the case of the intermediate scalar singlet
that couples to $(LL)_s$ inducing the effective interactions
in~(\ref{fierzed2}). Since ${\cal O}^e_\nu$ is not generated together
with ${\cal O}^e_\ell$ but only with operators that involve two
charged leptons and two neutrinos, the decay $\TauToMuee$ is of no use
to constrain the effective couplings.  However, since the effective
operators in~(\ref{fierzed2}) contain only left-handed fermions rather
strong bounds can be derived on the flavor diagonal terms using lepton
universality. The reason is that the corresponding interactions are
identical to the SM ones and have to be added coherently.

Setting $\alpha = \beta = \mu, \tau$ the last term in~(\ref{fierzed2})
induces additional contributions to $\ell_\alpha \to e_L \,
\nu_\alpha \, \lbar{\nu_e}$ with the effective coupling 
\beq
G_{\nu_\alpha \nu_\alpha}^e = 
 {|\lambda_{\alpha e}|^2 \over 4 \sqrt2 M^2} \,.
\eeq
These new contributions violate lepton universality and lead to a
deviation of the parameter
\beq \label{Rtaumu}
R_{\tau/\mu} \equiv \sqrt{{1 \over N}
{\Gamma(\tau^- \to e^- \nu_\tau \lbar{\nu_e}) \over
 \Gamma(\mu^- \to e^- \nu_\mu \lbar{\nu_e})}}  
\approx 
1 + {G_{\nu_\tau \nu_\tau}^e - G_{\nu_\mu \nu_\mu}^e \over G_F} 
\eeq
from unity. Here $N$ denotes a normalization factor, which is just the
ratio of the above two rates in the SM such that $R_{\tau/\mu} = 1$ if
$G_{\nu_\alpha \nu_\alpha}^e=0$. In the approximation we assume that
$G_{\nu_\alpha \nu_\alpha}^e \ll G_F~(\alpha=\mu, \tau)$. From the
most recent experimental data~\cite{Pich,PDG} it follows that
\beq 
R_{\tau/\mu} = 1.0008 \pm 0.0030 \,,
\eeq 
implying that
\beq \label{epbound} 
{\epsilon'}_\nu^e = 
 {G_{\nu_\tau \nu_\tau}^e - G_{\nu_\mu \nu_\mu}^e \over G_F} < 
 3.8 \times 10^{-3} \,.  
\eeq
Here we used that ${\cal O}_\nu^e$ has the same effective coupling as
the related operator that induces the new contribution to $\ell_\alpha
\to e_L \, \nu_\alpha \, \lbar{\nu_e}$.

Finally we remark that we can use lepton universality violation not
only to constrain the interactions induced by an intermediate singlet,
but our argument holds also whenever an $SU(2)_L$ related operator
induces additional contributions to the SM weak interactions. The
bounds on lepton universality violation in semi-hadronic
processes~\cite{Pich} are of similar order as the bounds for the
leptonic processes that appear in the definition of $R_{\tau/\mu}$
in~(\ref{Rtaumu}).  Consequently analogous arguments as those leading
to the upper bound on ${\epsilon'}_\nu^e$ in~(\ref{epbound}) can be
used to constrain ${\epsilon'}_\nu^q$.  Since all involved fermions
have to be left-handed this only applies for intermediate singlets or
triplets of $SU(2)_L$.  For the triplets the effective couplings of
the relevant operators may differ due to $SU(2)_L$ breaking effects,
which we will study next. 


\subsection{Constraining $SU(2)_L$ breaking effects}

If $SU(2)_L$ breaking effects are negligible then
$G_{\nu_\alpha\nu_\beta}^f$ is equal to $G_{\alpha\beta}^f$.
Comparing the experimental bounds (\ref{GeBound}), (\ref{GqBound}),
(\ref{GqBoundEta}) and (\ref{epbound}) with~(\ref{epsrequired}), we
find that in the $SU(2)_L$ symmetric limit the new neutrino
interactions that we considered cannot have a significant contribution
to the AN anomaly.

The excellent agreement between the SM predictions and the electroweak
precision data implies that $SU(2)_L$ breaking effects cannot be
large.  To show that they cannot sufficiently weaken the upper bounds
on $\epsilon_\nu^f$ and ${\epsilon_\nu'}^f$ to be consistent with
(\ref{epsrequired}), we recall from Eq.~(\ref{Gratio}) that in general
the ratio of the couplings, $G_{\nu_\alpha \nu_\beta}^f / G_{\alpha
\beta}^f$, is given by ratio $M_1^2 / M_2^2$.  Here $M_1$ and $M_2$
are the masses of the particles belonging to the $SU(2)_L$ multiplet
that mediate the processes described by $G_{\alpha \beta}^f$ and
$G_{\nu_\alpha \nu_\beta}^f$, respectively.  If $M_1 \ne M_2$ this
multiplet will contribute to the oblique parameters~\cite{obl} $S, U$
and, most importantly, $T$\,.  Then we can use a fit to the precision
data to determine the maximally allowed ratio $(M_1/M_2)^2_{\rm max}$.

We use the program {\tt GAPP} by J.~Erler~\cite{GAPP} to calculate the
SM predictions. For the latest precision data from the $Z$-pole
\cite{Z-pole}, the $W$-boson and top quark
masses~\cite{W-mass,t-mass}, deep inelastic scattering~\cite{DIS},
neutrino-electron scattering~\cite{nue} and atomic parity
violation~\cite{APV}, we obtain the best fit values of the oblique
parameters:
\beq
S = -0.07 \pm 0.11 \,, \qquad 
T = -0.10 \pm 0.14 \,, \qquad 
U = 0.11 \pm 0.15 \,.
\eeq
We calculate the contributions to $S$, $T$, and $U$ from the various
scalar representations in Tab.~1 and Tab.~2 and determine the best fit
to the data at each value of the mass splitting. The best fit Higgs
mass $M_H$ varies with the splitting, and we limit the Higgs mass to
the range 95~GeV $< M_H <$ 1~TeV.  Constructing a $\chi^2$ function we
determine the upper bound on the mass splitting between the different
members of a multiplet at a given confidence level (CL).  The
individual $90\%\,$CL bounds on $(M_1/M_2)^2$ are given in the last
column of Tab.~1 and Tab.~2. (Note that the limit on $(M_1/M_2)^2$ is
stronger if the lightest mass is heavier. From the $Z$-width
measurement, the lightest mass must be heavier than $M_Z/2$, and the
bounds presented in the tables are derived for the case where the
lighter mass is 50 GeV.)

We did not calculate the bounds for the vector multiplets. Since
vector bosons give in general larger contributions, we expect the
bounds in the vector cases to be as good or better than the
corresponding bounds for the scalar multiplets. Thus, for the vector
multiplets a rather conservative upper bound is $(M_1/M_2)^2 < 7$.

Hence, even the maximal possible $SU(2)_L$ breaking effects could
weaken the bounds we derived only by a factor of a few and
$\epsilon_\nu^f, {\epsilon'}_\nu^f$ cannot exceed the few-percent
level.  We learn that also the flavor changing neutrino scattering
\FCNS\ induced by heavy bosons, that are doublets or triplets of
$SU(2)_L$, cannot significantly contribute to the AN anomaly.

\section{$Z$-induced flavor changing neutral currents}
\label{Z-FCNC}

In the previous section we considered models where the only
modification to the neutrino sector is due to new interactions
mediated by heavy bosons.  In this section we study the opposite
scenario, where new fermions are added, but no extra bosons beyond the
SM ones are needed. As an example we consider $Z$-induced FCNCs that
arise when introducing a heavy sterile neutrino. Such SM gauge
singlets, which appear in many extensions of the standard model, are
frequently employed to explain the smallness of the neutrino masses
via the see-saw mechanism~\cite{seesaw}.

The basic idea for $Z$-induced FCNCs is that if a neutrino interaction
eigenstate is a linear combination of light and heavy mass eigenstates
then the effective low-energy interaction eigenstates, that consist
only of light mass eigenstates, are not orthogonal to each
other~\cite{non-seq}. Thus the couplings to the $Z$-boson (and in fact
also to the $W$-boson) have to be modified slightly, implying that
also the effective Hamiltonian that describes the neutrino propagation
in matter has to be changed.

In Ref.~\cite{BergmannKagan} a general discussion of $Z$-induced FCNCs
and their impact on neutrino oscillations has been presented.  In the
context of the AN problem we illustrate the mechanism by considering a
simple framework where besides the SM neutrinos $\nu_\mu$ and
$\nu_\tau$ there is only one new gauge singlet $\nu_S$.  (For
simplicity we assume that the $\nu_e$ does not play an important role
here.)  These interaction eigenstates are connected to the mass
eigenstates by a unitary transformation
\beq \label{mixing}
\pmatrix{\nu_\mu \cr  \nu_\tau \cr  \nu_S} =
\pmatrix{U_{\mu 1}  & U_{\mu 2}  & U_{\mu h}  \cr
         U_{\tau 1} & U_{\tau 2} & U_{\tau h} \cr
         U_{S1}     & U_{S2}     & U_{Sh} }
\pmatrix{\nu_1 \cr  \nu_2 \cr  \nu_h} \,,
\eeq
where $\nu_1$ and $\nu_2$ denote the light and $\nu_h$ the heavy mass
eigenstates.  The neutrinos that are produced in low-energy
charged-current interactions together with charged leptons $\mu$ and
$\tau$ are
\beq \label{effmix}
\pmatrix{\nu_\mu^P \cr  \nu_\tau^P} =
\pmatrix{U_{\mu 1}  & U_{\mu 2}  \cr
         U_{\tau 1} & U_{\tau 2} }
\pmatrix{\nu_1 \cr  \nu_2},
\eeq
i.e. we have projected $\nu_\mu$ and $\nu_\tau$ onto the $\nu_1-\nu_2$
plane.  Since the mixing matrix appearing in~(\ref{effmix}) is only a
submatrix of the unitary matrix in~(\ref{mixing}), $\nu_\mu^P$ and
$\nu_\tau^P$ are not orthogonal to each other
\beq
\bigr< \nu_\mu^P | \nu_\tau^P \bigr> = 
U_{\mu1}^* U_{\tau1} + U_{\mu2}^* U_{\tau2} = - U_{\mu h}^* U_{\tau h} 
\label{ortho}
\eeq
and also not properly normalized. Consequently these states do not
provide a proper basis for the neutrino oscillation formalism.

The description of neutrino oscillation in the presence of heavy gauge
singlets has been worked out in Ref.~\cite{BergmannKagan}. The main
result is that the effective non-unitary mixing induces a flavor
off-diagonal contribution in the matter-induced neutrino potential
$V_{FCNC}$. The effect is proportional to the neutron density and its
size is characterized by the ratio between $V_{FCNC}$ and the standard
(flavor diagonal) neutral current (NC) potential $V_{NC}$:
\beq \label{epsZ}
\epsilon_\nu^Z = {V_{FCNC} \over V_{NC}} \simeq |U^*_{\mu h} U_{\tau h}|
\eeq
The approximation refers to the simple example we discussed
previously.  It reveals that the effect is in general small, since it
is proportional to the components of the known neutrinos along the
heavy mass-eigenstates, which cannot be large.  The $Z$-induced FCNCs
cannot be constrained by the $SU(2)_L$-related charged lepton decay,
that we used before in the context of FCNIs due to heavy particle
exchange, but one can obtain a stringent constraint on
$\epsilon_\nu^Z$ from a global fit using lepton universality, CKM
unitarity, and the measured $Z$ invisible decay
width~\cite{limits}. The updated analysis in~\cite{BergmannKagan}
yields the conservative bounds (at 90\% CL)
\beq
|U_{\mu h}|^2 < 0.0096~~~~\mbox{and}~~~~|U_{\tau h}|^2 < 0.016 \,.
\label{constraint2} 
\eeq
We conclude that the parameter $\epsilon_\nu^Z$ in (\ref{epsZ}) cannot
exceed the few-percent level. Therefore from~(\ref{epsrequired}) it
follows that $Z$-induced FCNC effects are too small to be relevant for
the AN problem.

\section{Conclusions}
\label{conclusions}

Extensions of the standard model in general do not conserve lepton
flavor and therefore provide an alternative mechanism for neutrino
flavor conversion that may show up in neutrino oscillation
experiments. While such a scenario where flavor changing neutrino
interactions (FCNIs) explain one of the three neutrino anomalies is a
priori well motivated, one has to check carefully whether these
solutions are phenomenologically viable.  In~\cite{BergmannGrossman}
it was shown that FCNIs cannot be large enough to explain the LSND
anomalies. In this paper we argue that it is also very unlikely that
the AN anomaly is due to FCNIs. Both analysis rely on three facts:
\begin{itemize}
\item The neutrino flavor changing four fermion operator that is
induced by the exchange of a heavy boson is related by an $SU(2)_L$
rotation to other operators that violate lepton flavor.
\item The strength of these related operators is severely constrained
by the upper bounds on lepton number violating processes.
\item High precision measurements imply that the violation of the
$SU(2)_L$ symmetry is not large for new physics at or above the weak
scale. Consequently the upper bounds on the operators that induce
FCNIs are of the same order as those of the $SU(2)_L$ related
operators.
\end{itemize}
The first point follows immediately from the fact that the SM
neutrinos appear in $SU(2)_L$ doublets.  Using the upper bounds on
$\TauToMuM$ and $\TauToMuee$ we constrain, in a model-independent way,
the strength of the flavor changing neutrino scattering reaction
\FCNS\ to be at most at the one-percent level (compared with $G_F$).
For the unique case of an intermediate scalar singlet we derive a
severe constraint on the non-universal flavor diagonal neutrino
interactions using the upper bound on lepton universality violation.

The constraints we obtained for the parameters that describe the new
neutrino interactions are not consistent with the values that are
required to explain the AN anomaly in terms of FCNIs. Thus we conclude
that such a solution is ruled out. One could evade our bounds by
fine-tuning several new physics contributions, but we do not consider
such a scenario as very attractive. Nevertheless, we would like to
stress that ultimately any alternative explanation~\cite{Pakvasa} of
the AN anomaly should be tested by the experimental data itself. For
the time being ``standard'' neutrino oscillations with massive
neutrinos remain the most plausible and elegant solution.

\acknowledgements
 
We thank K.S. Babu, J. Erler, Y. Nir and S. Pakvasa for helpful
discussions.  Y.G. is supported by the U.S. Department of Energy under
contract DE-AC03-76SF00515. D.M.P. is supported under Department of
Energy contract DE-AC02-98CH10886.



\end{document}